# PARTITION SORT REVISITED: RECONFIRMING THE ROBUSTNESS IN AVERAGE CASE AND MUCH MORE!

Niraj Kumar Singh[1], Mita Pal[2] and Soubhik Chakraborty[3*]

[1]Department of Computer Science & Engineering, B.I.T. Mesra, Ranchi-835215, India
[2,3]Department of Applied Mathematics, B.I.T. Mesra, Ranchi-835215, India

```
*email address of the corresponding author:
       soubhikc@yahoo.co.in (S. Chakraborty)
```

## ABSTRACT

*In our previous work there was some indication that Partition Sort could be having a more robust average case O(nlogn) complexity than the popular Quick sort. In our first study in this paper, we reconfirm this through computer experiments for inputs from Cauchy distribution for which expectation theoretically does not exist. Additionally, the algorithm is found to be sensitive to parameters of the input probability distribution demanding further investigation on parameterized complexity. The results on this algorithm for Binomial inputs in our second study are very encouraging in that direction.*

## 1. Introduction

Average complexity is an important field of study in algorithm analysis as it explains how certain algorithms with bad worst case complexity perform better on the average like Quick sort. The danger in making such a claim often lies in not verifying the robustness of the average complexity in question. Average complexity is theoretically obtained by applying mathematical expectation to the dominant operation or the dominant region in the code. One problem is: for a complex code it is not easy to identify the dominant operation. This problem can be resolved by replacing the count based mathematical bound by a weight based statistical bound that also permits collective consideration of all operations and then estimate it by directly working on time, regarding the time consumed by an operation as its weight. A bigger problem is that the probability distribution over which expectation is taken may not be realistic over the domain of the problem. Algorithm books derive these expectations for uniform probability inputs. Nothing is stated explicitly that the results will hold even for non-uniform inputs nor is there any indication as to how realistic the uniform input is over the domain of the problem. The rejection of Knuth's proof in [1] and Hoare's proof in [2] for non uniform inputs should be a curtain raiser in that direction. Similarly,

                                                                                           23

it appears from [3] that the average complexity in Schoor's matrix multiplication algorithm is not the expected number of multiplications $O(d_1 d_2 n^3)$, $d_1$ and $d_2$ being the density (fraction of non zero elements) of pre and post factor matrices, but the exact number of comparisons which is $n^2$ provided there are sufficient zeroes and surprisingly we don't need a sparse matrix to get an empirical $O(n^2)$ complexity! This result is obtained using a statistical bound estimate and shows that multiplication need not be the dominant operation in every matrix multiplication algorithm under certain input conditions.

In our previous work [4] we introduced Partition Sort and found it to be having a more robust average case O(nlogn) complexity than the popular Quick sort. In our first study in this paper, we reconfirm this through computer experiments for inputs from Cauchy distribution for which expectation theoretically does not exist! Additionally, the algorithm is found to be sensitive to parameters of the input probability distribution demanding further investigation on parameterized complexity on this algorithm. This is confirmed for Binomial inputs in our second study.

**The Algorithm Partition Sort**

Partition-sort algorithm is based on divide and conquer paradigm. The function "partition" is the key sub-routine of this algorithm. The nature of partition function is such that when applied on input A[1.......n] it divides this list into two halves of sizes floor (n/2) and ceiling (n/2) respectively. The property of the elements in these halves is such that the value of each element in first half is less than the value of every element in the second half. The Partition-sort routine is called on each half recursively to finally obtain a sorted sequence of data as required. Partition Sort was introduced by Singh and Chakraborty [4] who obtained $O(nlog_2^2 n)$ worst case count, $\Omega(nlog_2 n)$ best case count and empirical $O(nlog_2 n)$ as the statistical bound estimate by working directly on time, for reasons stated earlier, in the average case.

## 2. Statistical Analysis

### 2.1 Reconfirming the robustness of average complexity of Partition Sort

**Theorem 1**: If U1 and U2 are two independent uniform U [0, 1] variates then Z1 and Z1 defined below are two independent Standard Normal variates:

Z1= $(-2\ln U1)^{1/2}$ Cos(2ЛU2); Z2= $(-2\ln U1)^{1/2}$ Sin(2ЛU2)

This result is called Box Muller transformation.

**Theorem 2**: If Z1 and Z2 are two independent standard Normal variates then Z1/Z2 is a standard

Cauchy variate. For more details, we refer to [5].

Cauchy distribution is an unconventional distribution for which expectation does not exist theoretically. Hence it is not possible to know the average case complexity theoretically for inputs from this distribution. Working directly on time, using computer experiments, we have obtained an empirical O(nlogn) complexity in average sorting time for Partition sort for Cauchy distribution inputs which we simulated using theorems 1 and 2 given above. This result goes a long way in reconfirming that Partition Sort's average complexity is more robust compared to that of Quick Sort. In [4] we have theoretically proved that its worst case complexity is also much better than that of Quick Sort as $O(n\log_2^2 n) < O(n^2)$. Although Partition Sort is inferior to Heap Sort's $O(n\log n)$ complexity in worst case, it is still easier to program Partition Sort.

Table 1 and figure 1 based on table 1 summarize our results.

Table 1: Average time for Partition Sort for Cauchy distribution inputs

| N | 10000 | 20000 | 30000 | 40000 | 50000 | 60000 | 70000 | 80000 | 90000 | 100000 |
|---|---|---|---|---|---|---|---|---|---|---|
| Mean Time (Sec.) | 0.05168 | 0.10816 | 0.1487 | 0.17218 | 0.20494 | 0.24078 | 0.2659 | 0.31322 | 0.35128 | 0.39496 |

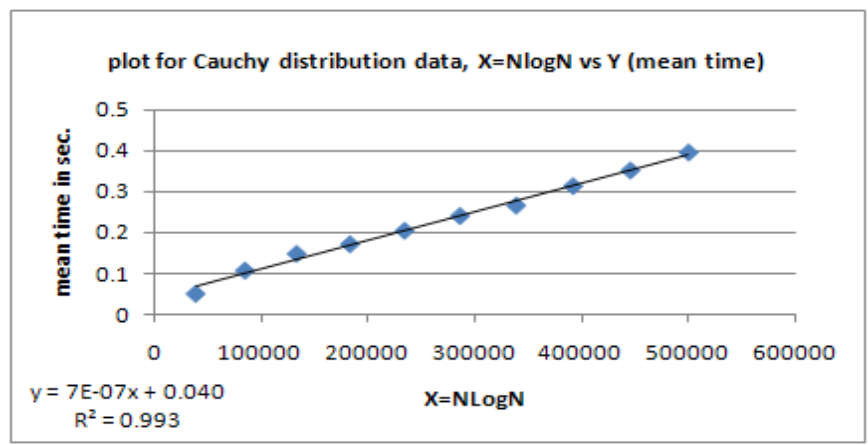

Fig. 1: Regression model suggesting empirical $O(n\log n)$ complexity

## 2.2 Partition Sort subjected to parameterized complexity analysis

**Parameterized complexity** is a branch of computational complexity theory in computer science that focuses on classifying computational problems according to their inherent difficulty with respect to *multiple* parameters of the input. The complexity of a problem is then measured as a function in those parameters. This allows to classify NP-hard problems on a finer scale than in the classical setting, where the complexity of a problem is only measured by the number of bits in the input. The first systematic work on parameterized complexity was done by Downey & Fellows [6]. The authors in [7] have strongly argued both theoretically and experimentally why for certain algorithms like sorting, the parameters of the input distribution should also be taken into account for explaining the complexity, not just the parameter characterizing the size of the input. The second study is accordingly devoted to parameterized complexity analysis whereby the sorting elements of Partition Sort come independently from a Binomial (m, p) distribution. Use is made of factorial experiments to investigate the individual effect of number of sorting elements (n), binomial distribution parameters (m and p which give the number of independent trials and the fixed probability of success in a single trial respectively) and also their interaction effects. A 3-cube factorial experiment is conducted with three levels of each of the three factors n, m and p. All the three factors are found to be significant both individually and interactively.

In our second study, Table-2 gives the data for factorial experiments to accomplish our study on parameterized complexity. For clarity, table 2 is presented in three parts- table 2.1, 2.2 and 2.3.

## Table 2: Data for $3^3$ factorial experiment for Partition Sort

Partition sort times in second Binomial ( m , p ) distribution input for various n (50000, 100000, 150000) , m ( 100 , 1000, 1500) and p (0.2, 0.5, 0.8).

Each reading is averaged over 50 readings.

Table 2.1 data for n = 50000

| m | p=0.2 | p=0.5 | p=0.8 |
|---|---|---|---|
| 100 | 0.07248 | 0.07968 | 0.07314 |
| 1000 | 0.09662 | 0.10186 | 0.09884 |
| 1500 | 0.10032 | 0.10618 | 0.10212 |

Table 2.2 data for n=100000

| M | p=0.2 | p=0.5 | p=0.8 |
|---|---|---|---|
| 100 | 0.16502 | 0.1734 | 0.16638 |
| 1000 | 0.21394 | 0.22318 | 0.21468 |
| 1500 | 0.22194 | 0.23084 | 0.22356 |

Table 2.3 data for n = 150000

| m | p=0.2 | p=0.5 | p=0.8 |
|---|---|---|---|
| 100 | 0.26242 | 0.27632 | 0.26322 |
| 1000 | 0.33988 | 0.35744 | 0.34436 |
| 1500 | 0.35648 | 0.37 | 0.35572 |

Table-3 gives the results using MINITAB statistical package version 15.

## Table-3: Results of $3^3$ factorial experiment on partition-sort

### General Linear Model: y versus n, m, p

```
Factor   Type    Levels  Values
n        fixed        3  1, 2, 3
m        fixed        3  1, 2, 3
p        fixed        3  1, 2, 3

Analysis of Variance for y, using Adjusted SS for Tests

Source  DF    Seq SS    Adj SS    Adj MS            F      P
n        2  0.731167  0.731167  0.365584  17077435.80  0.000
m        2  0.056680  0.056680  0.028340   1323846.78  0.000
p        2  0.001440  0.001440  0.000720     33637.34  0.000
n*m      4  0.011331  0.011331  0.002833    132322.02  0.000
n*p      4  0.000283  0.000283  0.000071      3302.87  0.000
m*p      4  0.000034  0.000034  0.000009       397.33  0.000
n*m*p    8  0.000046  0.000046  0.000006       266.70  0.000
Error   54  0.000001  0.000001  0.000000
Total   80  0.800982

S = 0.000146313   R-Sq = 100.00%   R-Sq(adj) = 100.00%
```

## 3. Discussion and more statistical analysis

Partition- sort is highly affected by the main effects n, m and p. When we consider the interaction effects, interestingly we find that all interactions are significant in Partition-Sort. Strikingly, even the three factor interaction n*m*p cannot be neglected. This means Partition Sort is quite sensitive to parameters of the input distribution and hence qualifies to be a potential candidate for deep investigation in parameterized complexity both theoretically (through counts) and experimentally (through weights) for inputs from other distributions. Further, we have obtained some interesting patterns showing how the Binomial parameters influence the average sorting time. Our investigations are ongoing for a theoretical justification for the same. The final results are summarized in tables 4-5 and figures 2A, 2B and 3 based on these tables respectively.
Each entry in the following tables is averaged over 50 readings.

Table 4: Partition Sort, Binomial (m, p) distribution, array size N=50000, p=0.5 fixed

| m | 100 | 300 | 500 | 700 | 900 | 1100 | 1300 | 1500 |
|---|---|---|---|---|---|---|---|---|
| Mean time (sec.) | 0.07968 | 0.09066 | 0.09586 | 0.09968 | 0.10154 | 0.10438 | 0.10282 | 0.10618 |

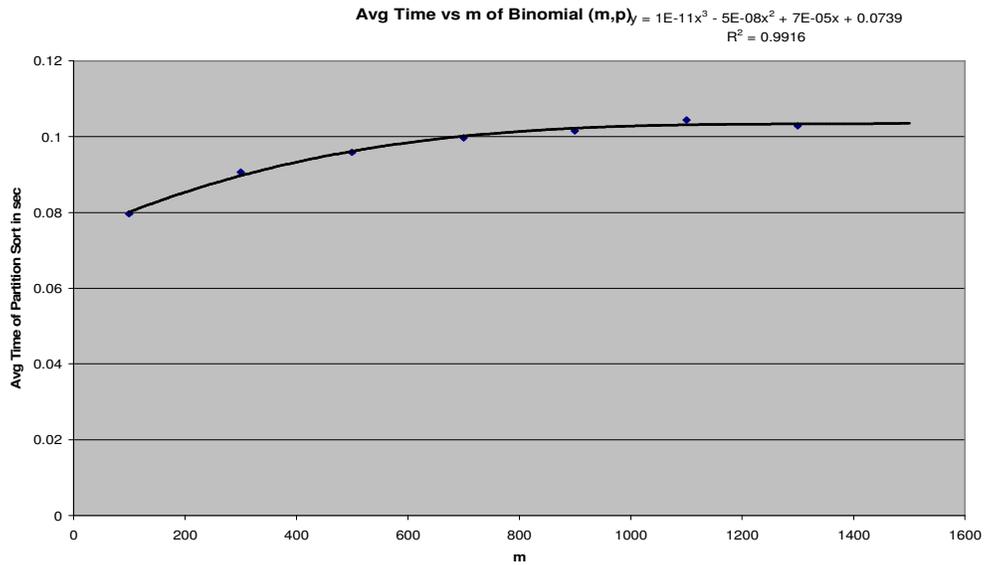

Fig 2A Third degree polynomial fit captures the trend

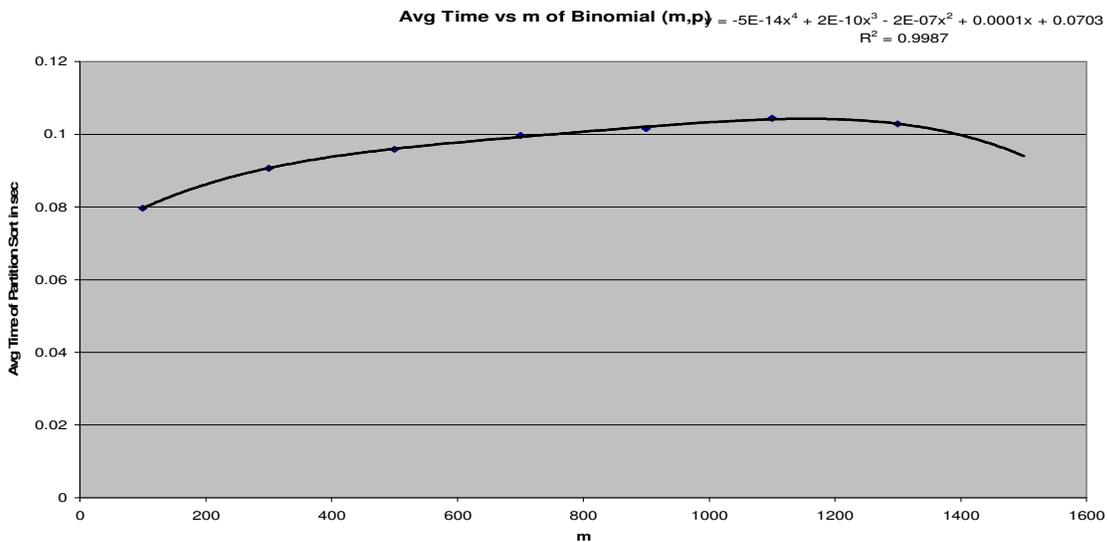

Fig 2B Fourth degree polynomial appears to be a forced fit (over fit)
(don't get carried away by the higher value of $R^2$!)

Although the fourth degree polynomial fit gives a higher value of $R^2$, it forces the fit to pass through all the data points. The essence of curve fitting lies in catching the trend (in the population) exhibited by the observations rather than catching the observations themselves (which reflect only a sample). Besides, a bound estimate must look like a bound estimate and it is stronger to write $y_{avg}(n, m, p) = O_{emp}(m^3)$ than to write $y_{avg}(n, m, p) = O_{emp}(m^4)$ for fixed n and p.

So we agree to accept the first of the two propositions.

Table 5: Partition Sort, Binomial distribution (m, p), n=50000, m=1000 fixed

| p | 0.1 | 0.2 | 0.3 | 0.4 | 0.5 | 0.6 | 0.7 | 0.8 | 0.9 |
|---|---|---|---|---|---|---|---|---|---|
| Mean time (sec.) | 0.09084 | 0.09662 | 0.09884 | 0.10198 | 0.10186 | 0.10034 | 0.0989 | 0.09884 | 0.09096 |

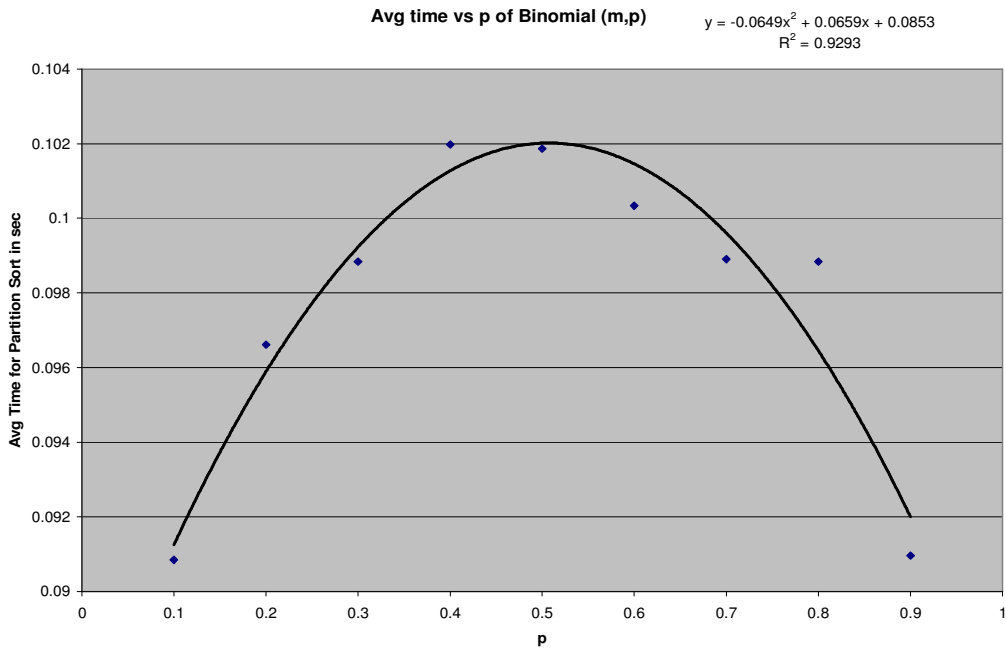

Fig. 3 Second degree polynomial fit captures the trend

Fitting higher polynomials lead to over-fitting (details omitted) and from previous arguments we put $y_{avg}(n, m, p) = O_{emp}(p^2)$ for fixed n and m.

For definitions of statistical bound and empirical O, we refer to [4]. For a list of properties of a statistical complexity bound as well as to understand what design and analysis of computer experiments mean when the response is a complexity such as time, [8] may be consulted.

## 4. Conclusion and suggestions for future work

We conclude

    (i)    Partition Sort is more robust than Quick Sort in average case.

International Journal of Computer Science, Engineering and Applications (IJCSEA) Vol.2, No.1, February 2012

(ii) Partition Sort is sensitive to parameters of input distribution also, apart from the parameter that characterizes the input size.

(iii) For n independent Binomial (m, p) inputs, all the three factors are significant both independently and interactively. All the two factor interactions nxm, nxp and mxp and even the three factor nxmxp is significant. This last finding is of paramount importance to excite other researchers on parameterized complexity and is intriguing if not impossible to be established theoretically. Theoretical analysis might confirm the influence of the Binomial parameters but how do you confirm the significance of their interactions? Using computer experiments where cheap and efficient prediction is the motive [8][9][10], we have settled the imbroglio.

(iv) We have also found $y_{avg}(n, m, p) = O_{emp}(m^3)$ for fixed n and p while $y_{avg}(n, m, p) = O_{emp}(p^2)$ for fixed n and m. It should be kept in mind that these results are obtained by working on weights and should not be confused with count based theoretical analysis which need not be identical.

In summary, this paper should convince the reader about the existence of weight based statistical bounds that can be empirically estimated by merging the quantum of literature in computer experiments (this literature includes factorial experiments, applied regression analysis and exploratory data analysis, which we have used here, not to speak of other areas like spatial statistics, bootstrapping, optimality design and even Bayesian analysis!) with that in algorithm theory. Computer scientists will hopefully not throw away our statistical findings and will seriously think about the prospects of building a weight based science *theoretically* to explain algorithm analysis given that the current count based science is quite saturated. This was essentially the central focus in our adventures. So the purpose achieved, we close the paper.